\documentclass[twocolumn,showpacs]{revtex4}

\usepackage{graphicx}%
\usepackage{dcolumn}
\usepackage{amsmath}
\newcommand{\be}{\begin{equation}}
\newcommand{\ee}{\end{equation}}
\newcommand{\bey}{\begin{eqnarray}}
\newcommand{\eey}{\end{eqnarray}}
\newcommand{\bw}{\begin{widetext}}
\newcommand{\ew}{\end{widetext}}
\newcommand{\ww}{\widetilde}
\newcommand{\ov}{\overline}
\newcommand{\ra}{\rangle}
\newcommand{\la}{\langle}
\newcommand{\br}{ {\bf r} }
\newcommand{\bp}{ {\bf p} }

\begin{document}

 \title {
 Uniform Semiclassical Approach to Fidelity Decay: from Weak to Strong Perturbation
 }

 \author{Wen-ge Wang$^{1,2}$ and Baowen Li$^1$}

 \affiliation{
 $^1$Department of Physics, National University of Singapore, 117542 Singapore
 \\ $^{2}$Department of Physics, Southeast University, Nanjing 210096, China
 }

 \date{02 November, 2004}

 \begin{abstract}

 We study fidelity decay by  a uniform semiclassical approach,
 in the three perturbation regimes, namely, the perturbative regime,
 the Fermi-golden-rule (FGR) regime, and the Lyapunov regime.
 A semiclassical expression is derived
 for fidelity of initial Gaussian wave packets with width of the order $\sqrt{\hbar }$
 ($\hbar$ being the effective Planck constant).
 Short time decay of fidelity of initial Gaussian wave packets is also studied,
 with respect to two time scales introduced in the semiclassical approach.
 In the perturbative regime, it is confirmed numerically that
 fidelity has the FGR decay before the Gaussian decay sets in.
 An explanation is suggested to a non-FGR decay in the FGR regime,
 which has been observed in a system with weak chaos in the classical limit,
 by using the Levy distribution as an approximation for the distribution of action difference.
 In the Lyapunov regime, it is shown that the average of the logarithm of fidelity may
 have roughly the Lyapunov decay within some time interval,
 in systems possessing large fluctuation of the finite-time Lyapunov exponent in the classical limit.

 \end{abstract}


\maketitle



 \section{Introduction}
 \label{sect:Intro}

 It is well known that, in classically chaotic systems, time evolution of trajectories
 in phase space is sensitive to small changes in
 initial conditions, as well as in parameters in the Hamiltonians.
 On the other hand, time evolution of state-vectors in Hilbert space
 is insensitive to small change of initial conditions.
 Nearly twenty years ago, Peres observed that
 small change in perturbation parameters can be employed in the study of the stability
 of quantum motion in the Hilbert space  \cite{Peres84},
 supported by further numerical investigations \cite{Peres91,SC96}.
 The quantity used to measure the stability of quantum motion
 is the quantum Loschmidt echo, or fidelity in the field of quantum information \cite{nc-book,gc-book}.
 It is an overlap of the evolution of the
 same initial state under two Hamiltonians with slight difference in the classical limit,
 $ M(t) = |m(t) |^2 $, where
 \be m(t) = \la \Psi_0|{\rm exp}(iH_1t/ \hbar ) {\rm exp}(-iH_0t / \hbar) |\Psi_0 \ra . \label{mat} \ee
 Here $H_0$ is the Hamiltonian of a classically chaotic system and
 $ H_1=H_0 + \epsilon V $  with $\epsilon $ being a small quantity.

 Fidelity decay has attracted increasing attention,
 since the work of Jalabert and Pastawski \cite{JP01},
 which relates the decay rate of fidelity to the (maximum) Lyapunov exponent of the
 underlying classical dynamics.
 In order to understand the behavior of fidelity in various systems,
 extensive investigations have been carried out
 \cite{ JSB01,JAB02,CPW02,CLMPV02,WVPC02,WC02,Prosen02,PZ02,CT02,CT03,BC02,EWLC02,WL02,STB03,Wis03,
 VH03,PSZ03,BCV03,AGM03,CDPZ03,CPJ03,WCL04,GPS03,WCLP04}.
 Previous investigations show the existence of at least four regimes
 of perturbation strength for fidelity decay \cite{CLMPV02, JAB02}:
 (i) In the perturbative regime below a perturbative border, fidelity has a Gaussian decay.
 In this regime, the typical transition matrix element is smaller than the mean level spacing.
 (ii) Above the perturbative regime is the Fermi-golden-rule (FGR)
 regime, with an exponential decay of the fidelity, $M(t) \propto {\rm exp}(-\Gamma t)$,
 where $\Gamma $ is the half-width of the local spectral density of states (LDOS) \cite{JSB01}.
 The decay rate can also be calculated semiclassically \cite{CT02}.
 (iii) With increasing perturbation strength, one enters into the Lyapunov regime,
 in which $M(t) \propto {\rm exp}(-\lambda t)$, with $\lambda $ being the Lyapunov exponent of
 the underlying classical dynamics \cite{JP01}.
 (iv) In the regime above the Lyapunov regime, perturbation strength is so large
 that the classical perturbation theory fails.
 Presently, little is known about the decaying behavior of fidelity in this regime.
 Within a random matrix theory approach, in which fidelity is expressed as the Fourier transform
 of LDOS, a Gaussian decay was suggested for fidelity decay in this regime \cite{JAB02},
 without further numerical confirmation (cf.~\cite{WIC98,CBH01} for shape of LDOS in this regime).
 It is also known that, for time $t$ short enough, fidelity has a quadratic decay,
 which may be extended to a Gaussian decay,
 just as a direct result of perturbation theory \cite{Wis03}.

 Most recent investigations show that the above picture of
 fidelity decay is incomplete, at least in four aspects.
 Firstly, in the perturbative regime, numerical results \cite{VH03} show that
 fidelity in the kicked rotator model has an exponential decay,
 which can be described by their semiclassical approximation,
 before the Gaussian decay sets in at about the Heisenberg time.
 A random matrix approach to fidelity also suggests an
 approximately exponential decay of fidelity at $t$ short enough \cite{GPS03}.
 It is not quite clear whether this exponential decay is the FGR decay or not.

 Secondly, a non-FGR decay of fidelity in the expected FGR regime has been found
 in a system with weak chaos in the classical limit,
 which is induced by deviation of the distribution of action difference from the
 expected Gaussian distribution \cite{WCL04}.
 An analytical description for the rate of the non-FGR decay is still lacking.

 Thirdly, in the Lyapunov regime, the decay rate of average fidelity
 has been found different from the Lyapunov exponent, although still  perturbation-independent,
 in systems possessing large fluctuation in the finite-time Lyapunov exponent \cite{Ott,ST02},
 as in the kicked top and  kicked rotator models \cite{JSB01,WL02,STB03}.
 A semiclassical WKB description of wave packets suggests an ${\rm exp}(-\lambda_1 t)$ decay
 for the fidelity, with $\lambda_1 < \lambda $ \cite{STB03}.
 More recently, more general semiclassical expressions of fidelity decay have been derived,
 with the Lyapunov decay and the $\lambda_1$ decay being two limiting cases \cite{WCLP04},
 along the lines of the semiclassical treatment to fidelity
 in Refs.~\cite{JP01,CT02,CLMPV02,VH03,WCL04}.
 However, the situation has not been clarified completely,
 since numerical results in the kicked top model show that
 the Lyapunov decay can be resumed in an approximate way,
 if average is performed on the logarithm of fidelity, but not on the fidelity itself \cite{WL02}.

 Finally, in the deep Lyapunov regime, fidelity of initial Gaussian wave packets
 may have a decay which is super-exponential and much faster than the Lyapunov decay at short initial times \cite{STB03}.
 Meanwhile, a decay with a rate of twice the Lyapunov exponent
 may appear before a time scale introduced in \cite{WCLP04},
 in systems with constant local Lyapunov exponents.
 A quantitative description of the former decay is still not available and
 the time scale that separates the two faster than Lyapunov decays is unknown

 In this paper, we use the uniform semiclassical approach introduced in \cite{VH03}
 to study the problems mentioned above.
 This approach is not only a suitable method for numerical evaluation of fidelity,
 but also a good starting point for analytical study \cite{WCL04,WCLP04}.
 For simplicity, we study 1D kicked systems only in this paper.

 The paper is organized as the following. In Sect.~\ref{sect:models}, we introduce
 two models, the kicked rotator and the
 sawtooth map, that will be employed for numerical check of our analytical results.
 A major difference between the two models is that
 the sawtooth map has a constant finite-time Lyapunov exponent,
 while the kicked rotator has large fluctuation in the finite-time Lyapunov exponent.
 For the sawtooth map,
 the semiclassical prediction of the rate of FGR-decay can be calculated accurately  at some parameter values,
 meanwhile, it has weak chaos with a structure of cantori in some parameter regime \cite{BCRHL99}.

 The validity of the uniform semiclassical approach
 has been checked numerically for initial point sources \cite{VH03,WCL04}.
 For initial Gaussian wave packets, narrowness of the packets are assumed
 in deriving the semiclassical expressions of fidelity in \cite{JP01,VH03}.
 In Sect.~\ref{sect:theory}, we show that the expression in \cite{VH03} fails
 in describing the fidelity decay,
 when the width of the initial Gaussian wave packet is of the order $ \sqrt {\hbar }$,
 where $\hbar $ is the effective Planck constant in the 1D kicked systems studied here.
 By considering the second order term in the Taylor expansion of the action,
 we derive a modified expression, which works well for this kind of initial wave packets.

 In Sect.~\ref{sect:short-time}, we discuss short time behavior of fidelity.
 In particular, a time scale is introduced for fidelity decay of initial
 Gaussian wave packets, which separates the two faster than Lyapunov decays mentioned above,
 and an analytical expression is derived for fidelity before this time scale.
 Dependence of the first-kick decay of fidelity of initial point sources on perturbation strength
 is also derived.
 Fidelity decay in the perturbative and the FGR regimes are studied in Sect.~\ref{sect:pert-FGR},
 with emphasis on the problems mentioned above.
 In the perturbative regime, the exponential fidelity decay before the Heisenberg time is
 shown numerically to coincide with the FGR decay in the sawtooth map.
 In the FGR regime, we revisit the non-FGR decay in the sawtooth map found in \cite{WCL04},
 and show numerically that the central part of the distribution of action difference can be approximated
 by the Levy distribution, which can explain some properties of the non-FGR decay.
 Section \ref{sect:Lyapunov} is devoted to a study of fidelity decay in the Lyapunov regime,
 when the average is performed over the logarithm of fidelity.
 Conclusions and discussions are given in Sect.~\ref{sect:con}.

 \section{Models: kicked rotator and sawtooth map}
 \label{sect:models}

 The Hamiltonians of the two models employed in this paper are of the forms,
 \bey  H= \frac 12 {p^2} + V_{ k(s)}(r) \sum_{n=0}^{\infty } \delta(t - n T)
 \hspace{1.5cm}   {\rm with } \label{H-both}
 \\ V_{ k}(r)= K \cos r \hspace{2cm} {\rm  for \ kicked \ rotator,  \ and } \label{H-kr}
 \\ V_{ s}(r)= - K (r- \pi)^2/2 \hspace{1cm} {\rm ( for \ sawtooth \ map). } \label{H-sawtooth}
 \eey
 For simplicity, the period $T$ is set to be unit, $T=1$.
 Kicks are switched on at $t=n$, $n=0,1,2, \ldots$.
 The classical map describing the kicked rotator is the standard map,
 \bey \nonumber p_{n+1} = p_n +  K \sin (r_{n}) \ \  ({\rm mod} \ 2\pi ),
 \\  r_{n+1} = r_n + p_{n+1} \ \hspace{0.85cm}  ({\rm mod} \ 2\pi ).  \label{map-kr} \eey
 The sawtooth map is
 \bey \nonumber p_{n+1} = p_n +  K (r_n - \pi ) \ \  ({\rm mod} \ 2\pi )
 \\  r_{n+1} = r_n + p_{n+1} \ \hspace{0.9cm}  ({\rm mod} \ 2\pi ).  \label{map-sawtooth} \eey

 Equation (\ref{map-sawtooth}) can be rewritten in the matrix form,
 \be \label{matrix-kr}
 \left( \begin{array}{c} p_{n+1} \\ r_{n+1}-\pi \end{array} \right )
 = \left( \begin{array}{cc} 1 & K \\ 1 & K+1 \end{array} \right )
 \left( \begin{array}{c} p_{n} \\ r_{n}-\pi \end{array} \right ), \ee
 where the $2 \times 2$ constant matrix possesses
 two eigenvalues $1+(K \pm \sqrt{K^2+4K} ) /2$.
 At $K>0$, motion in the sawtooth map is completely chaotic,
 with the Lyapunov exponent $\lambda = \ln \{ (2+ K + [ ( 2+K)^2 -4 ]^{1/2} ) /2 \}$,
 given by the largest eigenvalue of the matrix.
 The finite-time Lyapunov exponent
 has the same value as the Lyapunov exponent $\lambda $ defined in the limit $t \to \infty $.
 On the other hand, the standard map, which is chaotic at $K$ larger than 6 or so,
 does not have a constant finite-time Lyapunov exponent, because the mapping matrix is a function of $r_n$.
 It is of interest to mention a recent result on the classical analog of fidelity,
 namely, for systems with more than one dimensional configuration space
 the classical fidelity has a decaying rate related to not only the maximum Lyapunov exponent,
 but also other positive Lyapunov exponents \cite{VP03}.

 The two classical systems are quantized  on a torus
 \cite{HB80-q-tori, FMR91,WB94,Haake}.
 In a system with 1D finite configuration space, $0 \le r < r_m$,
 and 1D finite momentum space, $0 \le p < p_m$,
 the effective Planck constant $h$ and
 the dimension $N$ of the Hilbert space has the relation
 \be \label{h} N h =r_m p_m. \ee
 In both models, we take $r_m = p_m = 2 \pi $,
 hence, $\hbar = 2\pi / N$.

 Floquet operators in the two quantized systems  have the form
 \be \label{U} U = \exp [-i {\hat p}^2 / (2 \hbar )]
 \exp [-i V_{ k(s)}( {\hat r}) / \hbar ]. \ee
 Eigenstates of $\hat{r} $ are denoted by $|j \ra $, and $\hat{r}|j\ra = j \hbar |j\ra $, with  $j=0,1,2, \ldots , N-1$.
 In this representation, elements of the operator $U$ are
 \be U_{j'j} = \frac 1{\sqrt{N}} \exp \left [
 i \frac { \pi (j'-j)^2}{N} - i \frac{N V_{k(s)}(r_j)}{2 \pi }
 -i \frac{\pi}{4} \right ]. \label{Ujj} \ee
 The evolution of states,
 $ \psi (t)= U^t \psi_0 $, is calculated numerically
 by the fast Fourier transform (FFT) method.

 The fidelity in Eq.~(\ref{mat}) involves two slightly different Hamiltonians,
 $H_0$ and $H_1=H_0+ \epsilon V$.
 In what follows, $H_0$ takes the form of $H$ in Eq.~(\ref{H-both}), and
 \be \label{V-pt} V= \frac 1K V_{ k(s)}(r) \sum_{n=0}^{\infty } \delta(t - n T), \ee
 except in Sect.~\ref{sect:short-t-II}.

 \section{Uniform semiclassical approach to fidelity}
 \label{sect:theory}

 \subsection{Approximation to fidelity with action expanded to the first order term}

 For the sake of completeness and convenience in presenting our results,
  we briefly recall the main results of the semiclassical approach to
 fidelity in Refs.~\cite{JP01,VH03} in this section.

 In the semiclassical approach, an initial state $\psi_0 (\br_0)$ in a $d$-dimensional configuration space
 is propagated by the semiclassical Van Vleck-Gutzwiller
 propagator,
 \be \psi_{\rm sc} (\br ; t)= \int d\br_0  K_{\rm sc}(\br , \br_0  ; t)
 \psi_0 (\br_0 ), \label{psi-t} \ee
 where
 $ K_{\rm sc}(\br , \br_0  ; t)= \sum_{s} K_s(\br , \br_0  ; t)$,
 with
 \be K_s(\br , \br_0  ; t) = \frac{C_s^{1/2} } {(2 \pi i \hbar )^{d/2}}
 {\rm exp } \left [ \frac {i}{\hbar } S_s(\br , \br_0  ; t) -
 \frac{i \pi }2 \mu_s \right ]. \label{Ks-exp} \ee
 Here, the label $s$  [more exactly $s(\br , \br_0 ; t)$]
 indicates classical trajectories
 starting from $\br_0 $ and ending at $\br $ in a time $t$,
 the action $S_s(\br , \br_0  ; t)$ is the time integral of the Lagrangian
 along the trajectory $s$,
 $ S_s(\br , \br_0  ; t)= \int_0^t dt' {\cal L }$,
 and $ C_s = | {\rm det}( \partial^2 S_s / \partial {r_{0i}} \partial r_j )| $.
 $\mu_s$ is the Maslov index counting the conjugate points.

 Consider an initial Gaussian wave packet
 centered at $\ww \br_0$, with dispersion $\xi $ and mean momentum $\ww \bp_0$,
 \be \label{Gauss-wp} \psi_0 (\br_0 ) = \left ( \frac 1 {\pi \xi ^2} \right )^{d/4}
 {\rm exp} \left [ \frac i{\hbar} \ww \bp_0 \cdot \br_0
 - \frac{(\br_0 - \ww \br_0 )^2}{2 \xi ^2} \right ].  \ee
 When $\xi  $ is small enough,
 within the effective domain of $\br_0$,
 $ S_s(\br , \br_0  ; t)$ can be expanded in the Taylor expansion
 with respect to the center $\ww \br_0$,
 \be   S_s(\br , \br_0  ; t) = S_s(\br , \ww \br_0 ; t)
  - (\br_0  - \ww \br_0) \cdot {\bf p}_s + \ldots , \label{S-expan}  \ee
 where
 \be  {\bf p}_s = - \left . \frac{\partial S_s(\br , \br_0  ; t)}{\partial \br_0}
 \right |_{\br_0 = \ww \br_0}   \label{ps} \ee
 is the initial momentum of the trajectory $s(\br , \ww \br_0 ; t)$.

 The semiclassical approximation to the fidelity amplitude $m(t)$ in Eq.~(\ref{mat}) is
 \be \label{mt-sc-0} m(t) \simeq \int d\br \left [ \psi^{H_1}_{\rm sc}(\br ;t) \right ]^*
 \psi^{H_0}_{\rm sc}(\br ;t), \ee
 where the two states are propagated by the two Hamiltonians $H_1$ and $H_0$, respectively.
 For quite small $\xi $, the expansion in Eq.~(\ref{S-expan}) can be truncated at the first order term.
 Then, by using Eqs.~(\ref{Ks-exp})-(\ref{ps}) the integration on the right hand side of Eq.~(\ref{psi-t}) is calculated.

 The amplitude $m(t)$ thus obtained is \cite{JP01},
 \bey \nonumber m(t) \simeq m_{\rm sc1}(t) \equiv \left ( \frac{ \xi ^2 }{ \pi \hbar^2 } \right )^{d/2}
 \hspace{2.5cm} \\ \times  \int  d{\br } \sum_s C_s
 {\rm exp} \left [ \frac i{\hbar } \Delta S_s({\br }, \ww \br_0 ; t)
 - \frac{ \xi ^2 }{ \hbar^2 } (\bp_s - \ww \bp_0 )^2  \right ],
 \label{mt-gauss-r0-1st} \eey
 where $\Delta S_s(\br , \ww \br_0 ; t)$ is the action difference
 for the two trajectories with the same label $s$ in the two systems $H_1$ and $H_0$.
 In the first order classical perturbation theory, the difference between the two trajectories $s$ is assumed negligible,
 \be \Delta S_s(\br , \ww \br_0 ; t) \simeq \epsilon  \int_0^t dt' V[{\bf r}(t')] \label{DS} \ee
 with $V$ evaluated along the trajectory.

 A simpler expression of $m(t)$ can be obtained by
 changing the variable $\br \to \bp_0$ \cite{VH03},
 \bey \nonumber  m_{\rm sc1}(t)  = \left ( \frac{ \xi ^2 }{ \pi \hbar^2 } \right )^{d/2}
 \hspace{3cm} \\ \times  \int  d{\bp_0}
 {\rm exp} \left [ \frac i{\hbar } \Delta S({\bp_0}, \ww \br_0 ; t)
 - \frac{ (\bp_0 - \ww \bp_0 )^2  }{ (\hbar / \xi  )^2 }  \right ],
 \label{mt-gauss-p0-1st} \eey
 where $\Delta S(\bp_0 , \ww \br_0 ; t)$ coincides with $\Delta S_s
 (\br , \ww \br_0 ; t)$ for the same trajectory $s$ with initial momentum $\bp_0$.
 The main contribution to the right hand side of Eq.~(\ref{mt-gauss-p0-1st}) comes from a window
 in the $\bp_0$ space, which is centered at $\ww \bp_0$ and has a size of the order $\hbar / \xi $.

 For a system with finite momentum space, Eq.~(\ref{mt-gauss-p0-1st}) is invalid for initial
 Gaussian wave packets that are wide in the momentum space.
 The extreme case is for initial point sources, $\la \br | \Psi_0  \ra =
 \sqrt{(2 \pi \hbar )^d / {\cal V }_p} \delta (\br -\br_0) $, with ${\cal V}_p$
 being the volume of the momentum space.
 In this case \cite{WCL04},
 \be m(t) \simeq m_{p}(\br_0,t) \equiv  \frac 1{{\cal V}_p }  \int  d{\bf p}_0
 {\rm exp} \left [ \frac i{\hbar} \Delta S({\bf p}_0, \br_0 ; t)  \right ].
 \label{mt-point-p0} \ee
 The above semiclassical expressions of $m(t)$ suggests introducing
 \be \sigma = \epsilon / \hbar , \ee
  as a quantum perturbation parameter.

 \subsection{Contribution to fidelity from the second order term in action expansion}

 Hereafter we restrict our discussions to 1-dimensional kicked systems.
 In deriving Eq.~(\ref{mt-gauss-p0-1st}) for fidelity of an initial Gaussian wave packet,
 the right hand side of Eq.~(\ref{S-expan}) is truncated at the first order term.
 Hence, Eq.~(\ref{mt-gauss-p0-1st}) is valid only when $\xi  ^2 \ll \hbar $,
 or $\kappa \gg 1$ with the parameter $\kappa $ defined by
 \be \kappa \equiv \hbar / \xi ^2, \label{alpha} \ee
 which has been confirmed in our numerical calculation.

 \begin{figure}
 \includegraphics[width=\columnwidth]{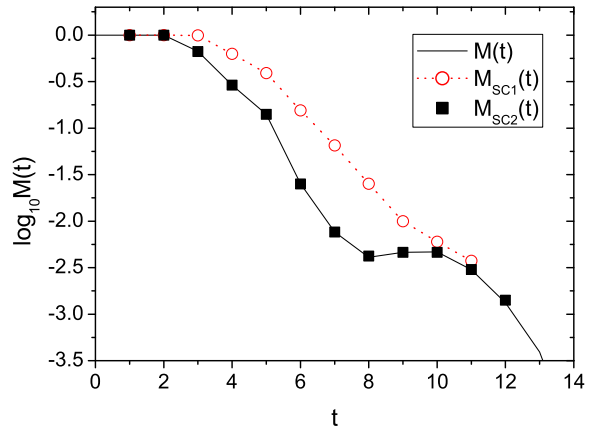}
 \caption{
 Comparison of the exact values of a single fidelity $M(t)$ and their semiclassical
 approximations $M_{\rm sc1}(t)$ in Eq.~(\ref{mt-gauss-p0-1st})
 and $M_{\rm sc2}(t)$ in Eq.~(\ref{mt-gauss-p0-2nd}),
 in the kicked rotator model. Parameters are  $K=10, N=2^{17}=131 072, \sigma =1$, and $\kappa =1$.
 The typical value of $D$ in Eq.~(\ref{D}) is about $8.9 \gg 1$,
 explaining the deviation of $M_{\rm sc1}(t)$ from the exact values,
 while $M_{\rm sc2}(t) = | m_{\rm sc2}(t)|^2 $ being quite close to the exact ones.
 } \label{fig-mt-sc-devia-tau1}
 \end{figure}

 When the condition $\kappa \gg 1$ is not satisfied,
 high order terms on the right hand side of Eq.~(\ref{S-expan})
 may give considerable contribution.
 Indeed, numerically, obvious deviation of $M_{\rm sc1}(t) = |m_{\rm sc1}(t) |^2$ from the exact $M(t)$
 has been observed  at $\kappa $ close to one or smaller
 (see Fig.~\ref{fig-mt-sc-devia-tau1} for an example).
 We remark that numerical evaluation of the right hand side of Eq.(\ref{mt-gauss-p0-1st}) for $m_{\rm sc1}(t)$
 becomes more and more difficult with increasing $t$,
 because the number of the oscillation of $\Delta S(p_0 , \ww r_0 ; t)$
 versus $p_0$ increases exponentially \cite{WCLP04}.

 To have a good semiclassical approximation at $\kappa \sim 1$,
 one needs to consider the second order term on the right hand
 side of Eq.~(\ref{S-expan}),
 \be  S_s(r , r_0  ; t) \simeq  S_s(r , \ww r_0 ; t)
  - (r_0  - \ww r_0) {p}_s - \frac 12  { \frac{\partial p_s}{\partial \ww r_0}}
  (r_0 - \ww r_0)^2 , \label{S-expan-2nd} \ee
 where
 \be   { \frac{\partial p_s}{\partial \ww r_0}}
\equiv \left . \frac{\partial p_s}{\partial r_0} \right |_{r_0 = \ww r_0 }
 = \left . - \frac{\partial^2 S_s (r,r_0 ; t)}{\partial r_0^2 }
 \right |_{r_0 = \ww r_0 }
  . \label{pp-pr0} \ee
 Using Eq.~(\ref{S-expan-2nd}) and following a procedure similar to
 the derivation of Eq.~(\ref{mt-gauss-p0-1st}),
 we obtain
 \bey \nonumber &   m_{\rm sc2}(t) =
 \hspace{3cm}
 \\ & \displaystyle   \int d{p_0} \frac {\xi  }{\sqrt{\pi } \ \hbar D }
 {\rm exp} \left [ \frac i{\hbar } \Delta S({p_0}, \ww r_0 ; t)
 - \frac{ (p_0 - \ww p_0 )^2 }{(\hbar D/ \xi  )^2  }  \right ],
 \label{mt-gauss-p0-2nd} \eey
 where
 \be D = \sqrt{ 1 + \frac 1{\kappa ^2}
 \left (  { \frac{\partial p_s}{\partial \ww r_0}} \right )^2 }. \label{D} \ee
 Note that $ D$ is  a function of $p_0,\ww r_0$ and $t$.
 A numerical test for this modified  semiclassical approximation is shown in Fig.~\ref{fig-mt-sc-devia-tau1}.

 Comparing the two equations (\ref{mt-gauss-p0-1st}) and (\ref{mt-gauss-p0-2nd}),
 it is seen that the modification is to replace  $(\hbar / \xi  )$
 in Eq.~(\ref{mt-gauss-p0-1st}) by
 \be w_p = \frac {\hbar }{\xi  }D = \sqrt{ \frac {\hbar ^2}{\xi ^2} + \xi ^2
 \left ( { \frac{\partial p_s}{ \ww \partial r_0}} \right )^2 }, \label{wp} \ee
 i.e., the change in the size of the effective window for integration.
 Therefore, the modified semiclassical expression in Eq.~(\ref{mt-gauss-p0-2nd})
 predicts the same long-time decaying behaviors of fidelity as Eq.~(\ref{mt-gauss-p0-1st}),
 more precisely, the same decaying rate for  the FGR decay in the FGR regime,
 and the same $\Lambda_1(t)$ decay in the Lyapunov regime (cf.~Sect.~\ref{sect:Lyapunov}).

 When the value of $\kappa  $ decreases further, higher order terms in
 the Taylor expansion of the action should be considered,
 e.g., at $\xi ^3 \sim \hbar $, the third order term should be taken into account.

 \section{Short time behavior of fidelity}
 \label{sect:short-time}

 The behavior of fidelity in short times is initial-state-dependent.
 In this section, we discuss short-time-decay of fidelity of
 initial Gaussian wave packets and of initial point sources.

 \subsection{Oscillation of $\Delta S(p_0,r_0;t)$ versus $p_0$}
 \label{sect:property-ds}

 The semiclassical expressions of fidelity discussed in the previous section,
 specifically, Eqs.~(\ref{mt-gauss-p0-1st}), (\ref{mt-point-p0}) and (\ref{mt-gauss-p0-2nd}),
 show that decaying behavior of $M(t)$ is mainly determined by
 the action difference $\Delta S(p_0,r_0 ;t)$ as a function of $p_0$.
 Therefore, before addressing fidelity decay, it is useful to first discuss an important property
 of the action difference, namely, its oscillation as a function of $p_0$.

 \begin{figure}
 \includegraphics[width=\columnwidth]{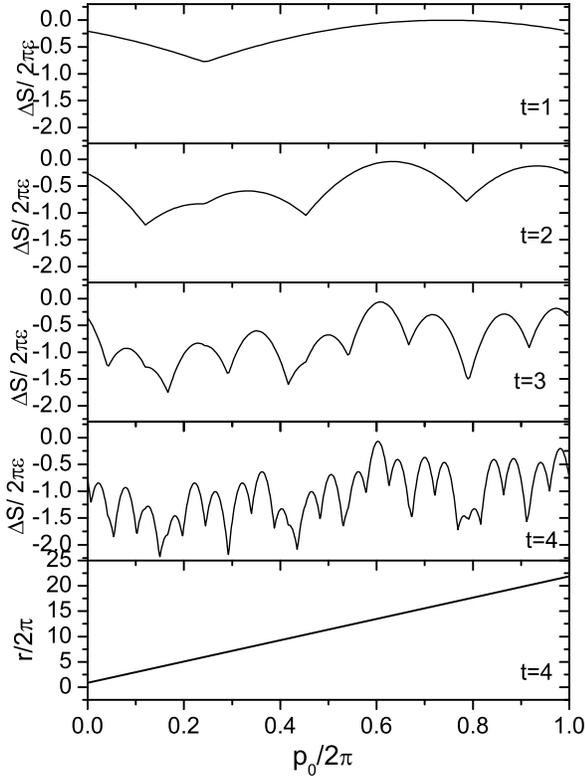}
 \vspace{0.2cm}
 \caption{
 Variation of $\Delta S(p_0,r_0;t)/ 2\pi \epsilon $ versus $p_0/2\pi $ in the sawtooth map
 at $K=1$, for a value of $r_0$ taken randomly within $[0,2\pi )$.
 $r(t)$ is the position $r$ at time $t$, with initial condition $(r_0,p_0)$.
 For clarity, $r$ is plotted as a continuous
 function of $p_0$, by adding  $2n\pi $ at the discontinuous points.
 } \label{fig-ds-p0-sawtooth}
 \end{figure}

 The number of the oscillations of $\Delta S$, as $p_0$ runs over $[0,2\pi )$,
 increases exponentially with time $t$.
 To see this, using Eq.~(\ref{DS}),
 we write the slope of $\Delta S / \epsilon $, denoted by $k_p$,
 in the following explicit form,
 \be k_p \equiv \frac{1}{\epsilon } \frac{ \partial  \Delta S (p_0,r_0 ;t)}{\partial p_0}
 \simeq  \int_0^t dt' \frac{\partial V}{\partial r'}
 \frac{\partial r'(t')}{\partial p_0}, \label{partial-ds} \ee
 where the dependence of $k_p$ on $p_0,r_0$ and $t$ is not
 written explicitly, for brevity.
 Due to the underlying  chaotic classical dynamics, $| \partial r'(t') / \partial p_0 |$
 increases exponentially with $t'$, on average.
 On the other hand, the variance of $\Delta S$ increases as $t$ \cite{CT02},
 hence, the typical value of $|\Delta S|$ increases as $\sqrt{t}$.
 Therefore, the number of the oscillations of $\Delta S$ must
 increase exponentially with $t$.
 The fast oscillation of $\Delta S$ is crucial for understanding
 the long time decay of fidelity in the Lyapunov regime  \cite{WCL04,WCLP04}.

\begin{figure}
 \includegraphics[width=\columnwidth]{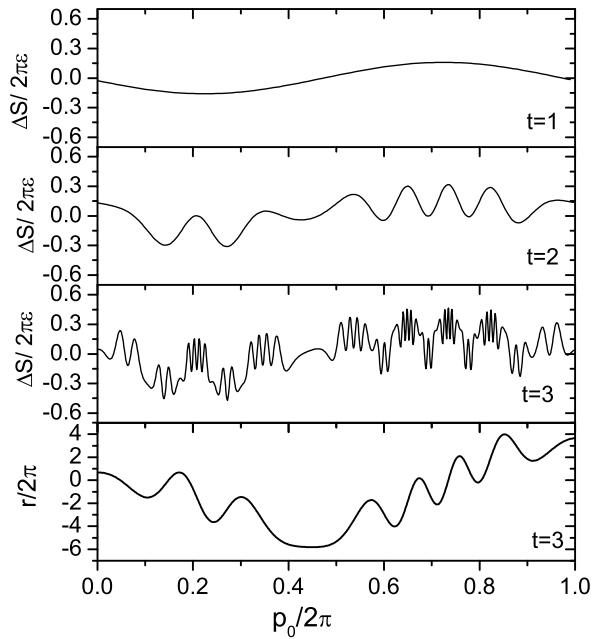}
 \caption{
 Same as Fig.~\ref{fig-ds-p0-sawtooth} but for the standard map of $K=10$.
 } \label{fig-ds-p0-kick-r}
 \end{figure}

 We present some examples of
 the oscillating behavior of $\Delta S$ in the sawtooth map (Fig.~\ref{fig-ds-p0-sawtooth}),
 as well as some in the kicked rotator model (Fig.~\ref{fig-ds-p0-kick-r}).
 The two models have different dependence of
 the position $r(t)$ on the initial momentum $p_0$ at fixed time $t$.
 Specifically, in the sawtooth map, $r$ is a linear function of $p_0$ except at the discontinuous points,
 with the slope given by the constant local Lyapunov exponent,
 while in the kicked rotator, it is an oscillating function.

 \subsection{Time scales $\tau_1$ and $\tau_2$ for fidelity of
 initial Gaussian wave packets}

 Fidelity of initial narrow Gaussian wave packets has rich behavior
 at short times.
 For example, there are both very fast and quite slow decays at the first several kicks
 in the deep Lyapunov regime \cite{STB03},
 as well as a decay with a rate of twice the Lyapunov exponent \cite{WCLP04}.
 By using the uniform semiclassical approach discussed above,
  we give a unified description for these phenomena in this section.

 \subsubsection{Time interval $t< \tau_1$}

 The main contribution to the right hand side of Eq.~(\ref{mt-gauss-p0-2nd})
 comes from a narrow window in the $p_0$ space.
 For time $t$ short enough, linear approximation can be used for the action difference $\Delta S$
 within the narrow window.
 This suggests the introduction of a time scale, denoted by $\tau_1$,
 such that for $t< \tau_1$ linear approximation of $\Delta S$ can be used
 in calculating the right hand side of Eq.~(\ref{mt-gauss-p0-2nd}),
 \be  \Delta S(p_0,\ww r_0 ; t)
 \simeq  \Delta S(\ww p_0,\ww r_0 ; t) + \epsilon \ww k_p (p_0- \ww p_0), \label{ds-small-t} \ee
 where $\ww k_p$ is the value of $k_p$ in Eq.~(\ref{partial-ds})
 at the center $(\ww r_0, \ww p_0)$ of the initial Gaussian packet.

 To give an estimation to $\tau_1$, we use $\Delta p_0(t)$ to denote
 the size of the region in the $p_0$ space, which is capable of the above linear approximation for $\Delta S$.
 One should note that $\Delta p_0(t)$ shrinks exponentially,
 due to the exponentially increasing of the number of oscillations of $\Delta S$ versus $p_0$.
 Since the oscillation of $\Delta S$ is mainly induced by local instability of trajectories,
 $\Delta p_0(t)$ shrinks roughly as $ e^{-\Lambda(t) t} $, where
 \be \\  \Lambda (t)  = \lim_{\delta x(0) \to 0}  \frac 1t  \overline{ \ \left [ \ln \left |
 \delta x(t)/ \delta x(0) \right |  \right ] \ }, \label{Lamb-t}  \ee
 with $\delta x(t)$ denoting distance in phase space and average performed over phase space.
 (In a classical system with strong chaos, $\Lambda (t)$ usually approaches the
 Lyapunov exponent $\lambda $ quickly, as will be illustrated numerically in Sect.~\ref{sect:Lyapunov}.)
 Then,
 \be \label{dp0t} \Delta p_0(t) \simeq b(t) \Delta p_0(1) e^{-\Lambda (t) \cdot (t-1)}, \ee
 where $b(t)$ is the influence of other factors, such as the variance of $\Delta S$ increasing
 linearly with t, and changes much slower than the exponential term.
 At time $\tau_1$, we write
 \be \label{t1-condi} \Delta p_0(\tau_1) = a_1 w_p , \ee
 where $a_1>1$  is determined by the accuracy required.

 Substituting Eq.~(\ref{dp0t}) into Eq.~(\ref{t1-condi}) for $t=\tau_1$, we obtain
 \be \tau_1 \sim 1+ \frac 1{\Lambda (\tau_1)} \ln  \frac{\ov b \Delta p_0(1) }{a_1 w_p}, \label{t1} \ee
 where $\ov b$ is the average value of $b(t)$ for small $t$.
 Several points can be seen in Eq.~(\ref{t1}).
 Firstly, since $\Delta p_0(1)$ decreases with increasing $\sigma $, when $\sigma $ is large enough,
 the right hand side of Eq.~(\ref{t1}) can be smaller than 1, implying $\tau_1=0$.
 Secondly, for large enough $\tau_1$ such that $\Lambda (\tau_1) \simeq \lambda$,
 the dependence of $\tau_1$ on $\hbar $ is given by
 $(1/2\lambda ) \ln \hbar ^{-1} $ for $\xi  = \hbar ^{1/2}$, which is half the Ehrenfest time.

 When the change of $D$ is negligible within the effective narrow window of $p_0$,
 in which the linear approximation for $\Delta S$ in Eq.~(\ref{ds-small-t}) can be used,
 substituting Eq.~(\ref{ds-small-t}) into Eq.~(\ref{mt-gauss-p0-2nd}), we have
 \be M_{\rm sc2}(t) \simeq \exp \left [- \frac 12 \left (  {\sigma w_p \ww k_p}
 \right )^2 \right ], \hspace{1cm} t<\tau_1, \label{Mt-gauss-short-t} \ee
 with the time dependence on the right hand side given by $\ww k_p (\ww r_0, \ww p_0 ; t)$.

 Due to exponential divergence of neighboring trajectories in phase space,
 the main contribution to the right hand side of Eq.~(\ref{partial-ds}) comes from times $t \approx t'$.
 Therefore, $|\ww k_p|$ increases typically as $ c_k e^{\Lambda (t)t}$, with $c_k$ being the pre-factor.
 For this time dependence of $|\ww k_p |$, Eq.~(\ref{Mt-gauss-short-t}) predicts
 \be M_{\rm sc2}(t) \simeq \exp \left [- \frac 12 w_p^2 c_k^2
 \left (  \epsilon / \hbar \right )^2 e^{2\Lambda (t) t} \right ],
 \hspace{0.4cm} t<\tau_1. \label{Mt-gauss-sh-t-est} \ee
 This gives the extremely fast, double-exponential-rate decay of fidelity
 predicted in Ref.~\cite{STB03}, $ M(t) \simeq \exp ( -\text{const}\times e^{2\lambda t} ) $,
 when $\Lambda (t) \sim \lambda $.
 Note that our prediction (\ref{Mt-gauss-short-t}) is more general, since
 it works for small values of $|\ww k_p|$ as well, so long as $\ww p_0$ is not
 quite close to any stationary point, which may invalid the approximation in Eq.~(\ref{ds-small-t}).

\begin{figure}
 \includegraphics[width=\columnwidth]{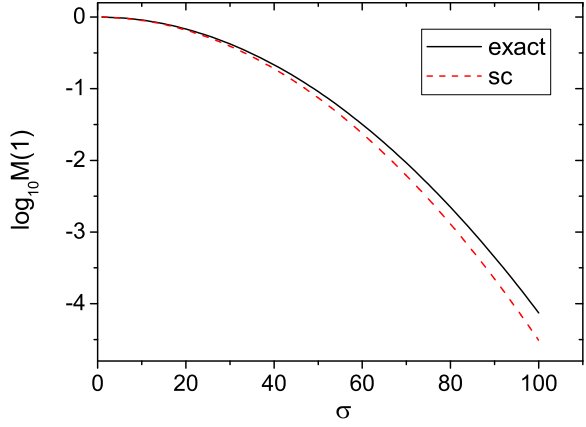}
 \caption{
 Comparison of the exact values of $M(t)$ and the semiclassical prediction given by Eq.~(\ref{Mt-gauss-short-t})
 at the first kick $t=1$, for one initial Gaussian wave packet,
 in the kicked rotator model with $K=10$, $\kappa =1$.
 } \label{fig-short-t-V1st}
 \end{figure}

 Numerical check of our prediction (\ref{Mt-gauss-short-t})
 is shown in Fig.~\ref{fig-short-t-V1st} for the first kick.
 For $\sigma < 40$, the analytical results have good agreement with the exact numerical calculations.
 With increasing $\sigma $, deviation enlarges, with $M_{sc}(t) < M(t)$,
 because the difference between the exact values of the phase $\Delta S / \hbar $
 and their linear approximations increases linearly with $\sigma $.

 \subsubsection{Time interval $\tau_1 < t < \tau_2$}
 \label{sect:short-t-II}

 For $t> \tau_1$, the main contribution to the right hand side of
 Eq.~(\ref{mt-gauss-p0-2nd}) comes from the integration over the region $[\ww p_0 - w_p,\ww p_0 + w_p]$.
 It is useful to introduce a second time scale, denoted as $\tau_2$, at which
 $\Delta S(p_0,\ww r_0;t)$ completes one full oscillation period as
 $p_0$ runs over $[\ww p_0 - w_p,\ww p_0 + w_p]$ \cite{WCLP04}.
 Note that $\tau_1 < \tau_2$, according to their definitions.

 In order to estimate the time $\tau_2 $,
 we note that the number of oscillations of $\Delta S$ increases as $e^{\Lambda (t) t}$,
 then, $\tau_2$ satisfies the following relation
 \be \label{t2-gene} \ov \tau_2  \approx \frac{1}{\Lambda (\ov \tau_2)} \ln \left ( \frac {\pi }{c_0 w_p } \right ). \ee
 When $\Lambda (\ov \tau_2) \simeq \lambda $, this relation gives the estimation
 \be \label{t2} \ov \tau_2  \approx \frac{1}{\lambda } \ln \left ( \frac {\pi }{c_0 w_p } \right ). \ee


 The time scale $\tau_2$ is important in understanding short-time decay of fidelity
 in the deep Lyapunov regime with $\sigma \gg 1$.
 Indeed, in the time interval $\tau_1 < t < \tau_2$,
 the phase $(\Delta S/ \hbar )$ on the right hand side of Eq.~(\ref{mt-gauss-p0-2nd}),
 as a function of $p_0$, can usually be approximated by a straight line
 within the region $p_0 \in [\ww p_0 - w_p,\ww p_0 + w_p]$.
 Then, for initial states satisfying $|\sigma \ww k_{p}| \gg \pi / w_p $, one has \cite{WCLP04},
 \be M_{\rm sc}(t) \propto 1/ (\sigma \ww k_{p})^2, \hspace{1.cm} \tau_1 <t<\tau_2  . \label{M-skp} \ee

 To be more specific, let us consider a special kind of system,
 which has constant local Lyapunov exponent $\lambda $ in the classical limit
 and has no stationary point of $\Delta S$, i.e.,  $k_p \ne 0$ for all $p_0$.
 For such systems, $\Lambda (t) =\lambda $.
 As shown in \cite{WCLP04}, when the smallest $|k_p|$ are so large that Eq.~(\ref{M-skp}) is applicable
 for all initial states,
 the average fidelity has a decay with a rate of twice the Lyapunov exponent,
 \be \ov M(t) \propto e^{-2\lambda t}, \label{M-2-lambda-t} \ee
 since $k_p$ increases as $e^{\lambda t}$ on average.
 Here we are interested in the general situation, in which the smallest $|k_p|$ are not large
 enough for the applications of Eq.~(\ref{M-skp}), i.e., when $|\sigma \ww k_{p}| \gg \pi / w_p $ is not satisfied
 for $|\ww k_p|$ close to its smallest value.
 In this general case, the average fidelity has a decay rate smaller than $2\lambda $ due to the influence the small $|k_p|$.
 On the other hand, we note that the size of the region of $p_0$ with quite small $|k_p|$
 decreases exponentially, due to the exponential increment of the number of oscillation of $\Delta S$,
 therefore, the decay rate of the average fidelity should be larger than $\lambda $.

 \begin{figure}
 \includegraphics[width=\columnwidth]{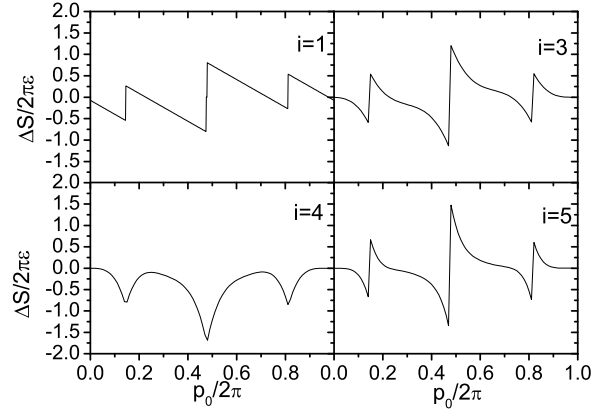}
 \caption{
 Same as Fig.~\ref{fig-ds-p0-sawtooth} but for the sawtooth map with perturbations
 in Eq.~(\ref{Vit}) and $t=2$.
 } \label{fig-ds-p0-Vi-1-5}
 \end{figure}

 To verify numerically the above prediction of fidelity decay between
 $e^{-\lambda t}$ and $e^{-2\lambda t}$ for $\tau_1 < t< \tau_2$,
 we use the sawtooth map, with the following form of the perturbation $V$
 ($H_0$ unchanged) \cite{WCLP04},
 \bey  V^{(i)}=  V^{(i)}(r) \sum_{n=0}^{\infty } \delta(t - n T)
 \hspace{1cm } {\rm where } \label{Vit}
 \\ V^{(i)}(r) = - {\cal N}_i (r - \pi )^i, \hspace{1cm}
 i=1,2,3,4,5. \label{Vi} \eey
 Setting the coefficient ${\cal N}_2 = 1/2$, $V^{(2)}$ gives the perturbation
 in Eq.~(\ref{V-pt}), which are also used in \cite{BC02,WCL04}.
 The other coefficients ${\cal N}_i$ are chosen by the requirement
 of having the same decaying rate in the FGR regime,
 i.e., possessing the same value of the classical action diffusion constant $K(E)$ \cite{CT02}.
 For kicked maps, $K(E)$ has the form \cite{LCT99,CLLT00},
 \bey \label{K-kick}  K(E) = \frac 12 C(0) + \sum_{l=1}^{\infty } C(l),
 \hspace{1cm} {\rm where }
 \\ \label{C(l)} C(l) = \left \la \{ V[r(l)] -\la V \ra \}
 \{  V[r(0)] - \la V \ra \} \right \ra  \eey
 with the average performed over phase space.

 \begin{figure}
 \includegraphics[width=\columnwidth]{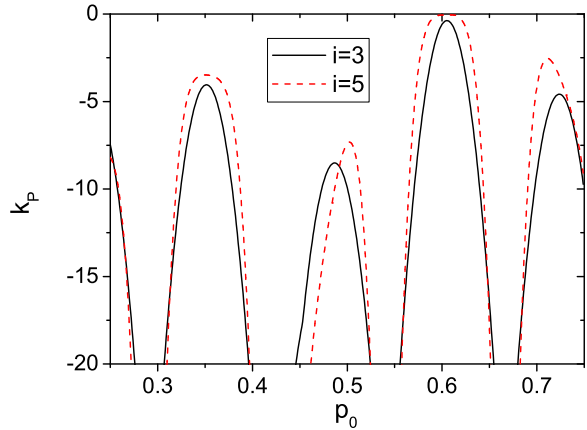}
 \caption{
 Values of $k_p$ of a $r_0$ chosen randomly, in the sawtooth map with perturbation
 $V^{(i)}$ of $i=3$ and 5, $t=3$.
 } \label{fig-kp-i3-i5-st}
 \end{figure}

 At integer values of $K$ in the sawtooth map in Eq.~(\ref{H-sawtooth}),
 simple derivation shows that $C(l) =0$ for $l \ne 0$ and
 \be \label{c0}  \displaystyle C(0) = \left \{ \begin{array}{l}
  \displaystyle \frac 1{2i+1} {\cal N}_i^2 \pi^{2i},
  \hspace{2cm} ({\rm odd} \ i);
 \\  \displaystyle \frac {i^2}{(2i+1)(i+1)^2}
 {\cal N}_i^2 \pi^{2i}, \hspace{1cm} ({\rm even} \ i).
 \end{array} \right .  \ee
 Then,
 \be \label{Ni} {\cal N}_1 = \frac {\pi }{\sqrt{15}}, \
 {\cal N}_3 = \frac {\sqrt {1.4} }{3 \pi }, \
 {\cal N}_4 = \frac {\sqrt{5}}{4 \pi^2}, \
 {\cal N}_5 = \frac { \sqrt{2.2} }{3 \pi^3 }.  \ee

 \begin{figure}
 \includegraphics[width=\columnwidth]{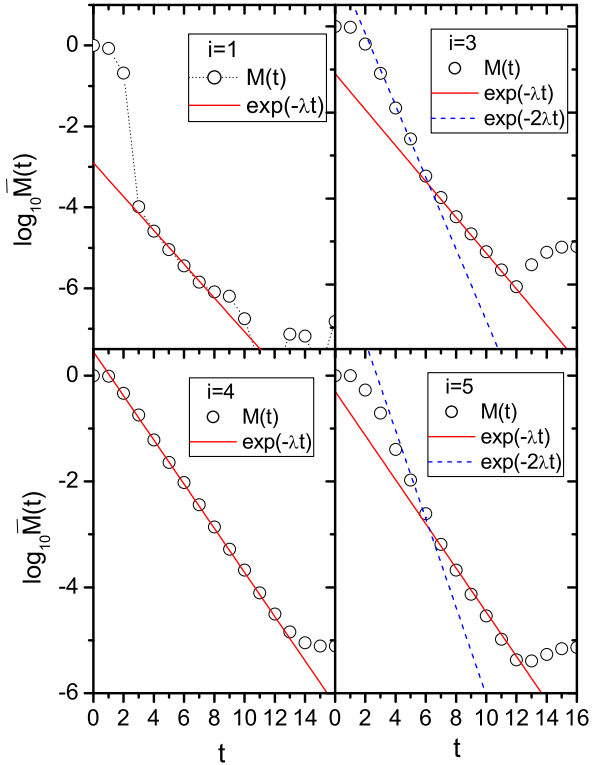}
 \vspace{0.0cm}
 \caption{
 Decay of the averaged fidelity in the sawtooth map with $K=1$,
 for $V^{(i)}$ of $i=1,3,4,5$ in Eq.~(\ref{Vit}),
 with parameters $\sigma =100$, $N=131072$, $\xi  = \sqrt{\hbar }$.
 For these parameters, $\ov \tau_1 \sim 1$ and $\ov \tau_2 \approx 6.5$.
 In the time interval $\tau_1 < t < \tau_2$,
 the average fidelity has the Lyapunov decay for $i=4$,
 a double-Lyapunov decay for $i=3$, and a decay between the two decays
 for $i=5$.
 The quite fast decay for $i=1$ is due to the linear dependence of
 $\Delta S$ on $p_0$, as shown in Fig.~\ref{fig-ds-p0-Vi-1-5}.
 Averages are performed over 2000 initial Gaussian packets,
 with centers taken randomly with flat distribution
 in the region $\pi /2 \le \ww r_0 (\ww p_0) < 3\pi /2 $.
 } \label{fig-Mt-vr1345-v1st-s100}
 \end{figure}

 For the sawtooth map with the above perturbation,
 the action difference at time $t$ can be written as
 \be \label{DS-st} \Delta S \simeq - \epsilon \sum_{n=0}^{t-1} {\cal N}_i
 [ r(n) - \pi ] ^i. \ee
 It is easy to prove, by using Eq.~(\ref{map-sawtooth}), that $r(n)$
 for any fixed $n$ is a monotonically
 increasing function of $p_0$, except at the discontinuous changes
 from $0$ to $2\pi $ (or reversely).
 Then, Eq.~(\ref{DS-st}) shows that no point exists at which
 $k_p =0$ for odd $i$,
 while $k_p$ can be zero for even $i$.
 (See Figs.~\ref{fig-ds-p0-sawtooth} and \ref{fig-ds-p0-Vi-1-5}  for some examples
 of numerical illustrations.)
 Therefore, in the time interval $\tau_1 < t < \tau_2$,
 $\ov M(t)$ should have the standard Lyapunov decay for even $i$ \cite{WCLP04}
 and faster than Lyapunov decay for odd $i$.

 Some values of $|k_p|$ for $V^{(i)}$ of $i=3$ and 5 are presented in Fig.~\ref{fig-kp-i3-i5-st}.
 It is seen that some $k_p$ for $i=5$ are quite close to zero, implying a decay rate of fidelity
 between $\lambda $ and $2\lambda $ in the time interval $(\tau_1 , \tau_2)$ for $i=5$;
 on the other hand, the smallest $|k_p|$ for $i=3$ is not quite close to zero,
 implying a decay rate of $2\lambda $.
 Indeed, these predictions have been confirmed in a direct calculation of $M(t)$,
 as shown in Fig.~\ref{fig-Mt-vr1345-v1st-s100}.
 The values of $\tau_1$ and $\tau_2$ can be estimated as follows.
 At $K=1$ and $\xi  = \sqrt{\hbar }$, numerical computation shows that $c_0 \simeq 0.45$ and $D \simeq 1.9$.
 We take $\Delta p_0(1) \sim 2\pi / 100$
 for $\sigma =100$ (cf.~Fig.~\ref{fig-ds-p0-sawtooth} for variation of $\Delta S / \epsilon $
 at $t=1$), $a_1 \sim 5$, $\ov b \sim 1$, then, Eq.~(\ref{t1}) gives $\ov \tau_1 \sim 1 $.
 Meanwhile, Eq.~(\ref{t2}) gives  $\ov  \tau_2 \approx 6.5$.
 The two estimations are in good agreement with the direct numerical results shown in Fig.~\ref{fig-Mt-vr1345-v1st-s100}.

 \subsection{Dependence of first-kick-decay of fidelity on perturbation strength for initial point sources}
 \label{sect:first-kick}

 \begin{figure}
 \includegraphics[width=\columnwidth]{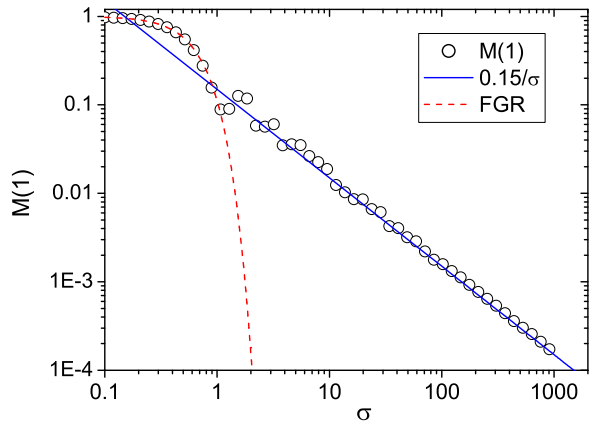}
 \caption{
 Decay of fidelity $M(t)$ at the first kick $t=1$ as a function of $\sigma $, for a single initial point source,
 in the logarithm scale.
 It shows a FGR behavior for small $\sigma $ and a $1/ \sigma $  dependence  for large $\sigma $.
 $M_{\rm FGR}(t) \simeq e^{-2.2 \sigma^2 t}$ in the sawtooth map with $K_0=1$.
 } \label{fig-m1-tau}
 \end{figure}

 Fidelity of initial point sources, described by Eq.~(\ref{mt-point-p0})
 with integration performed over the whole $p_0$ domain,
 has a short-time decay different from that of initial Gaussian wave packets discussed above.

 In the FGR regime, fidelity can be calculated by writing the right hand side of
 Eq.~(\ref{mt-point-p0}) in terms of the distribution of $\Delta S$ (see Refs.~\cite{CT02,WCL04}).
 When the distribution of $\Delta S$ is close to the expected Gaussian distribution,
 one has the FGR decay for fidelity,
 \be \label{Mt-fgr} M_{\rm FGR}(t) \simeq \exp [ - 2\sigma^2 K(E) t], \ee
 where $K(E)$ is the classical action diffusion constant in Eq.~(\ref{K-kick}).

 In the deep Lyapunov regime, Eq.~(\ref{mt-point-p0}) enables an estimation to
 the dependence of fidelity on perturbation strength,
 which is $1/\sigma $, as shown in Eq.~(\ref{Mt-p-tau}) in Appendix \ref{sect:proof-1st-kick}.
 Since this dependence does not change with time, it can be seen at the first kick only.
 For systems with constant local Lyapunov exponents,
 combining Eq.~(\ref{Mt-p-tau}) and the known Lyapunov decay, we have
 \be \ov M_p(t) \propto \frac 1{\sigma } \exp (-\lambda t).  \label{Mt-p-tau-lamb} \ee
 Figure \ref{fig-m1-tau} presents an example of numerical confirmation to the above predictions
 for the first-kick decay of fidelity.

 \section{The Perturbative and the FGR regimes}
 \label{sect:pert-FGR}

 In this Section, we study fidelity decay before the Heisenberg time
 in the perturbative regime,
 and the influence of weak chaos on fidelity decay in the FGR regime.

 \subsection{The perturbative regime}

 The regime of quite small $\epsilon $, more precisely, quite small $\sigma $,
 is named the perturbative (PT) regime,
 in which fidelity has a Gaussian type decay \cite{JSB01,CT02,CT03}.
 Combining the perturbation theory, the random matrix theory (RMT),
 and the semiclassical theory, it has been found that
 \be \label{Mt-pt-gaus}
 M_{\rm PT}(t) \simeq \exp \left ( - \frac{2g K(E) }{\pi \ov d \beta } \sigma^2 t^2 \right ), \ee
 for quantized maps,
 where $2g/ \beta $ is the number of classical orbits with identical action,
 which is 2 for the models used here,
 and $\ov d $ is the total mean density of states.
 The index $\beta =1$ for time-reversal-invariant systems and $\beta =2$ for
 time-reversal-breaking systems.
 (Cf.~Ref.~\cite{CT03} for the expression for continuous variables.)

 \begin{figure}
 \includegraphics[width=\columnwidth]{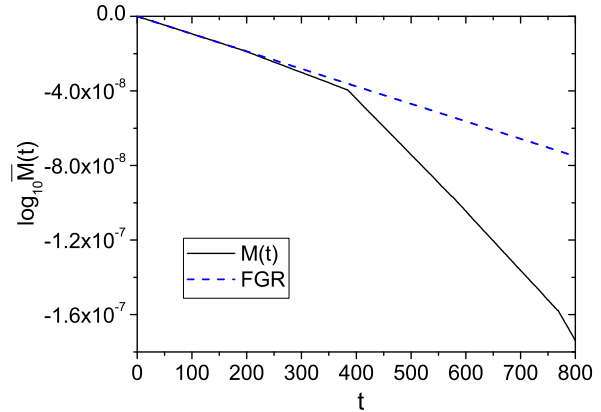}
 \caption{
 Decay of the averaged fidelity in the perturbative regime of the sawtooth map
 at $K=1$, $N=512$, and $\sigma =10^{-5}$.
 Average is performed over 100 initial point sources taken randomly in the
 configuration space.
 The decay follows the FGR decay approximately up to $t \sim 400$.
 } \label{fig-Mt-s1d-5-n9-st}
 \end{figure}

 The Gaussian decay (\ref{Mt-pt-gaus}) sets in at about the Heisenberg time $t_H \approx N$.
 For $t$ short enough compared with $t_H$,
 there is no reason for the semiclassical approach to fail in describing the fidelity
 decay, even for quite small $\sigma $.
 Therefore, it is reasonable to expect the FGR decay in Eq.~(\ref{Mt-fgr}) for fidelity
 in the perturbative regime, before some time shorter than $t_H$.
 In fact, numerically, fidelity has been found to possess an exponential decay in the kicked rotator model,
 before some time shorter than the Heisenberg time $t_H$ \cite{VH03}.
 Whether this exponential decay is the FGR decay or not is unclear,
 since the value of $K(E)$ in this model is calculated only approximately.

 To make the situation clearer, we employ the sawtooth map, in which the value of $K(E)$
 can be computed analytically for integer values of the parameter $K$ in its Hamiltonian,
 $K(E) = \pi^4 /90 \simeq 1.08 $ [see Eq.~(\ref{c0})].
 Numerical results indeed support the above argument that fidelity has the FGR decay,
 before a time shorter than the Heisenberg time (see Fig.~\ref{fig-Mt-s1d-5-n9-st}).

\begin{figure}
 \includegraphics[width=\columnwidth]{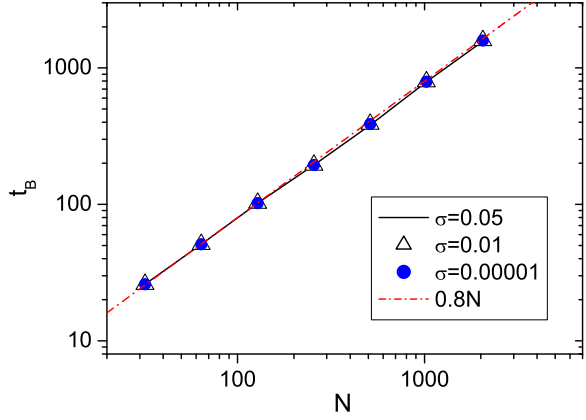}
 \caption{
 The breakdown time $t_B$ of the semiclassical approach versus the dimension $N$ of the Hilbert space,
 in the logarithmic scale for the sawtooth map
 with $K=1$, at different values of $\sigma $.
 In numerical calculation, $t_B$ is taken as the first kick at which
 the relative error $|[M(t)- M_{\rm FGR}(t)] / M(t)|$ is larger than 0.1,
 where $M_{\rm FGR}(t)$ is the FGR decay in Eq.~(\ref{Mt-fgr}) predicted by the
 semiclassical theory.
 } \label{fig-tb-n-st-point}
 \end{figure}

 Deviation of fidelity decay from the FGR decay in the perturbative regime,
 provides a good opportunity for
 a numerical study of the breakdown time of the semiclassical approach, denoted by $t_B$.
 It is known that $t_B$ is proportional to some inverse algebraic power of $\hbar $
 \cite{TH91,STH92,CTH92,TH03}.
 In the sawtooth map, it was found that $t_B$ is linear in $1/ \hbar $,
 more exactly, $t_B \approx 0.8N \propto \hbar^{-1}$, as seen in Fig.~\ref{fig-tb-n-st-point}.

 \subsection{The FGR regime}

 When the perturbation parameter $\sigma $ is large enough,
 $M(t)$ becomes close to its saturation value, which is proportional to $1/N$ \cite{PZ02},
 before the Gaussian decay sets in at about the Heisenberg time,
 then, one enters into the FGR regime.
 The perturbation border for the crossover from the perturbative regime to the FGR regime
 can be estimated as \cite{CT02}
 \be  \sigma_p = \frac{ \epsilon_p }{\hbar } \sim \sqrt{ \frac{\ln N }{ 2K(E)N}}. \ee

 For systems possessing strong chaos in the classical limit,
 two analytical approaches are available to obtain the FGR decay,
 namely, the semiclassical approach with the assumption of a Gaussian type distribution of $\Delta S$\cite{CT02},
 which gives the result in Eq.~(\ref{Mt-fgr}),
 and the RMT approach making use of the closeness between the form of the LDOS
 and the Lorentzian distribution \cite{JSB01,JAB02}.
 The two approaches are believed to be equivalent, when both are valid;
 while an analytical proof of the equivalence is available only in some special cases
 \cite{CLMPV02}.
 An interesting phenomenon is that the two approaches are complementary in some cases.
 For example,  for quite short time $t$, there is no analytical reason for the
 distribution of $\Delta S$ to be close to a Gaussian distribution,
 while the RMT approach can be used in deriving the FGR decay.
 On the other hand, for $t$ relatively long (but shorter than $t_B$),
 the semiclassical approach works well,
 while the RMT approach may meet the problem of deviation of the LDOS from the Lorentzian form in the tail region,
 which is a result of the finite domain of the quasi-energy spectrum \cite{WL02}.

 When the underlying classical dynamics has weak chaos,
 non-FGR decay of fidelity may appear in the expected FGR regime, due to
 obvious deviation of the distribution of $\Delta S$ from the expected Gaussian distribution \cite{WCL04}.
 For this kind of systems, the RMT approach does not give a correct prediction for fidelity decay,
 while the semiclassical approach still works.

 In the semiclassical approach, one can separate the average fidelity into a mean-value part and a
 fluctuating part, denoted by  $\overline M_a(t) $ and $\overline M_f(t)$, respectively,
 $\overline M(t) \equiv \overline{|m(t)|^2} = \overline M_a(t) + \overline M_f(t)$,
 where
 \be \label{Mp} \overline M_a(t) \equiv |  \overline m(t) |^2,  \ee
 with average performed over initial states.
 In the FGR regime, the average fidelity is approximately given by
 the mean-value part $\ov M_a(t)$,
 with $\ov M_f(t) \ll \ov M_a(t)$ \cite{WCL04}.
 The mean-value part $\overline M_a(t)$ can be expressed in
 terms of the distribution $P(\Delta S)$ of the action difference $\Delta S$,
 \bey \overline M_a(t) \simeq \left | \int d\Delta S e^{i\Delta S/ \hbar }
 P(\Delta S)\right |^2, \hspace{0.2cm} {\rm where} \hspace{1.4cm}  \label{Mp-ps}
 \\ P(\Delta S) = \frac 1{ \int d\br_0 d{\bf p}_0 }
 \int d\br_0 d{\bf p}_0 \delta \left [\Delta S - \Delta S(\bp_0 , \br_0;t) \right ].
 \label{PdS}  \eey

\begin{figure}
\includegraphics[width=\columnwidth]{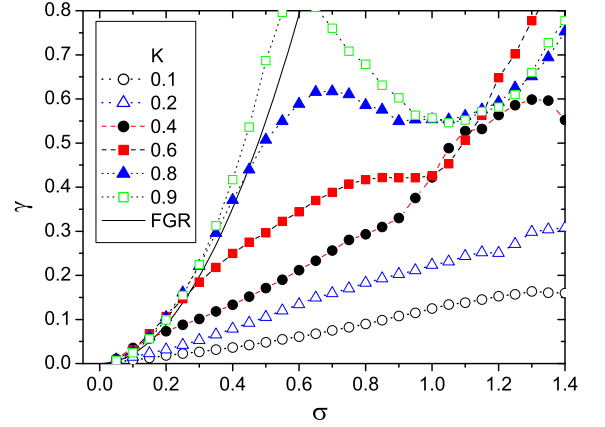}
 \caption{
 Decay rate $\gamma $ of fidelity versus perturbation
 strength $\sigma $, calculated by the best fit of $-\gamma t$ to $\ln \overline M(t)$,
 for some values of the parameter $K$ in the sawtooth map.
 In the calculation of average fidelity,
 1000 randomly chosen Gaussian wave packets are used as initial states.
 The solid curve shows the rate $\Gamma \simeq 2.2\sigma^2$ of the FGR-decay.
 $N=131072$, $\xi = \sqrt{\hbar }$.
 } \label{fig-gam-sig-k04}
\end{figure}

 In case of weak chaos, a general analytical expression for the distribution of $\Delta S$ is still absent.
 Since Gaussian distribution is invalid in this case,
 it is natural to study the possibility of Levy distribution.
 Due to its infinite variance, Levy distribution cannot describe
 the distribution of $\Delta S$ in the long tail region.
 Therefore, we focus on the central part and short tail region of the distribution of action difference,
 which give the main contribution to the mean-value part $\ov M_a(t)$ of fidelity.

 We consider the following asymmetric  form of  Levy distribution \cite{Umeno},
 \be L(x,\alpha , \beta ) = \frac 1{2\pi } \int_{-\infty }^{\infty }
 \exp (izx) \psi (z) dz, \label{Levy} \ee
 with $x=\Delta S / \epsilon $.
 Here the function $\psi (z)$ is
 \bey \psi (z) = \exp \{ -igz -D_l|z|^{\alpha } [ 1+i\beta \ {\rm sgn}(z) \ \omega (z,\alpha )] \}, \label{Levy-in}
 \\ \text{where} \hspace{1.7cm} \omega (z,\alpha ) = \tan (\pi \alpha /2) \ \ \text{for} \ \ \alpha \ne 1,
 \\ \text{     }  \hspace{1.7cm} \omega (z,\alpha ) = (2/\pi ) \ln |z| \ \ \text{for} \ \ \alpha = 1.\eey
 The parameter $\alpha $, with $0<\alpha < 2$, determines the decay of long tails,
 i.e., $L(x) \sim |x|^{-(1+\alpha )}$ for large $|x|$.
 The parameter $\beta $ has the domain $[-1,1]$, with $\beta =0$ giving the symmetric distribution,
 $g$ gives a shift along the $x$ direction, and $D_l$ is related to the width of the distribution.
 If the Levy distribution can be used as an approximation for $P(\Delta S)$,
 substituting Eq.~(\ref{Levy}) into Eq.~(\ref{Mp-ps}),
 one obtains,
 \be  \ov M_a(t)  \propto \exp ({- 2 D_l \ \sigma^{\alpha }}), \label{Mt-Levy} \ee
 with the time dependence given by that of $D_l$.
 Note that a Gaussian form of $P(\Delta S)$ corresponds to $\alpha =2$,
 giving the well known dependence on $\sigma $ in the FGR decay.

 It is known that the sawtooth map has weak chaos at $K<1$
 and has  Cantori structures at $K$ small \cite{BCRHL99}.
 Non-FGR decay has been observed in this model, which can still be described  by Eq.~(\ref{Mp-ps})  \cite{WCL04}.
 Therefore, we use this model to study the possibility of using the Levy distribution
 as an approximation to the distribution $P(\Delta S)$.

\begin{figure}
\includegraphics[width=\columnwidth]{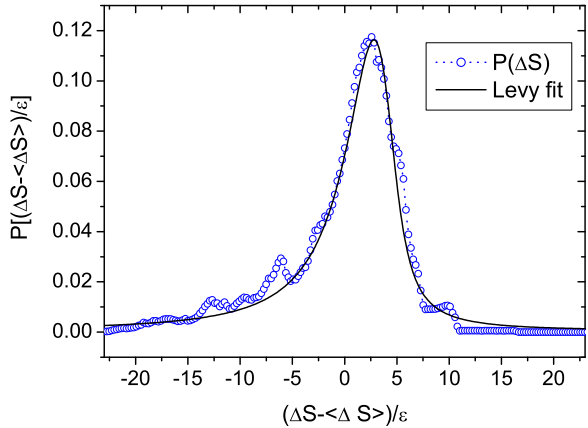}
 \caption{
 Distribution $P[(\Delta S -\la \Delta S \ra )/ \epsilon]$
 of the classical action difference $\Delta S$  at $t=10$, for $K=0.4$ in the sawtooth map,
 calculated by taking randomly $10^7$ initial points in the phase space.
 Here $\la \Delta S \ra \equiv \epsilon t \la V \ra = -\pi^2 \epsilon t/6$,
 with average performed over the phase space.
 The solid curve is a fit given by the Levy distribution in
 Eq.~(\ref{Levy}), with $\alpha =1$, and $\beta$ and $D_l$ as two fitting parameters.
 } \label{fig-ps-k04-levy}
\end{figure}

 Numerically we use ${-\gamma t}$ to fit $\ln \ov M(t)$, in order to calculate the decay rate $\gamma $ of fidelity.
 Variation of $\gamma $ versus $\sigma $ is presented in Fig.~\ref{fig-gam-sig-k04}
 for some values of $K$ between 0.1 and 0.9, before the Lyapunov regime is reached.
 (The fidelity has been found to have the Lyapunov decay in the Lyapunov regime,
 for $K<1$ \cite{WCL04}.)
 As seen in the Figure, for $K =0.1, 0.2$, and 0.4, there exist some regions of $\sigma $, respectively,
 in which $\gamma $ increases approximately linearly with $\sigma $.
 According to Eq.~(\ref{Mt-Levy}), this implies that $\alpha \simeq 1$, if the Levy
 distribution can be used as an approximation to $P(\Delta S)$.
 Figure \ref{fig-ps-k04-levy} shows a fit of the Levy distribution to the central part
 and short tails of the distribution $P(\Delta S)$ at $K=0.4$,
 with $\alpha = 1$ fixed and $D_l$ and $\beta $ used as two adjusting parameters.
 The agreement is encouraging, for which an analytical explanation is still not yet available and deserves a further investigation.

 \section{The Lyapunov regime }
 \label{sect:Lyapunov}

 Increasing the perturbation strength further, one enters into the Lyapunov regime,
 in which the average fidelity has the Lyapunov decay $e^{-\lambda t}$ \cite{JP01},
 in the special situation with negligible fluctuation in the finite-time Lyapunov exponent.
 In the general situation, the fluctuation of the finite-time Lyapunov exponent is not negligible,
 and the average fidelity has a $\Lambda_1(t)$ decay \cite{WCLP04}, which is usually different from the Lyapunov decay.
 The $\Lambda_1(t)$ decay will be discussed briefly in Sect.~\ref{sect:spa}, starting from
 the modified semiclassical approximation to fidelity in Eq.~(\ref{mt-gauss-p0-2nd}).

 As mentioned in the introduction, in the kicked top model that has large fluctuation in the
 finite-time Lyapunov exponent, numerical results  show that the average of the logarithm of fidelity
 has roughly the Lyapunov decay \cite{WL02}.
 In Sect.~\ref{sect:avg-ln}, we explain this phenomenon by using the technique developed in \cite{WCLP04}.

 \subsection{$\Lambda_1(t)$ decay of average fidelity in the deep Lyapunov regime}
 \label{sect:spa}

 Since Eq.~(\ref{mt-gauss-p0-2nd}) can be obtained from
 Eq.~(\ref{mt-gauss-p0-1st}) by replacing $\hbar / \xi $ with $w_p = \hbar D / \xi $,
 generalization of the results in \cite{WCLP04} is straightforward.
 In this Section, we present the main points in the generalization,
 because part of them will be used in Sect.~\ref{sect:avg-ln}.

 For systems in which $\Delta S$ has stationary points with $k_p=0$,
 we denote by $\alpha $ the stationary points of $\Delta S$ and by $p_{0\alpha }$
 the momenta at which $k_p=0$.
 When $\sigma \gg 1$, stationary phase approximation  can be used in calculating
 the right hand side of Eq.~(\ref{mt-gauss-p0-2nd}), which gives
 \bey \label{msc-ma} m_{\rm sc}(t) & \simeq & \sum_{\alpha } m_{\alpha }(t), \ \ \ \ \  \text{where} \
 \\ \label{ma-t} m_{\alpha }(t) & = & \frac{\sqrt{ 2 i \hbar }}{w_p }
 \frac{ {\rm exp} \left [ \frac i{\hbar} \Delta S( p_{0\alpha },\ww r_0 ; t)
 - (p_{0\alpha } - \ww p_0 )^2/w_p^2 \right ] }
 {\sqrt{ | \Delta {S_{\alpha }^{''}} | }}, \nonumber
 \\ & & \text{with} \ \ \ \  \Delta {S_{\alpha }^{''}} = \left .
 \frac{ \partial^2 \Delta S(p_0,\ww r_0;t) }{ \partial p_0^2 }
 \right |_{p_0= p_{0\alpha }}. \hspace{0.3cm}  \label{ma-st} \eey
 Note that $w_p$ in Eq.~(\ref{ma-t}) takes the value at the stationary
 point $\alpha $, with $D$ being a function of $p_{0\alpha }$.

 Let us consider time $t> \tau_2$, for which there are one or more stationary points within the
 effective window in the $p_0$ space for the integration on the right hand side of Eq.~(\ref{mt-gauss-p0-2nd}).
 Similar results can also be obtained for $t<\tau_2$, when the stationary phase approximation
 is applicable, as discussed in \cite{WCLP04}.
 The average fidelity in the ordinary sense, with average performed over both $\ww r_0$ and $\ww p_0$,
 can be calculated by using diagonal approximation, when $\sigma \gg 1$ \cite{WCL04},
 and the following result is obtained
 \be \label{Mt-It}  \ov M(t) \propto  I_s(t) := \int d \ww r_0 \int_{ {\cal P}_{\delta }} dp_0
 \frac 1{ D | k_p |}, \ee
 where ${\cal P}_{\delta } := \bigcup_{\alpha } {\cal A}_{\alpha }$.
 Here, ${\cal A}_{\alpha }$ denotes the region
 $[p_{0\alpha }^-,p_{0\alpha }-\delta ] \bigcup [p_{0\alpha }+ \delta ,p_{0\alpha}^+]$,
 where $p_{0\alpha}^{-}= ({p}_{0\alpha } +{p}_{0,\alpha -1})/2$,
 $p_{0\alpha}^{+}= ({p}_{0\alpha } +{p}_{0,\alpha +1})/2$,
 with $\delta $ being a small quantity.

 The main contribution to the integral on the right hand side of Eq.~(\ref{Mt-It})
 comes from small values of $|k_p|$ in the region ${\cal P}_{\delta }$.
 For $p_0 \in {\cal P}_{\delta }$ close to a stationary point $p_{0\alpha }$,
 $k_p$ in Eq. (\ref{partial-ds}) can be approximated by
 \be k_p \approx \int_0^t dt' \left [
 \frac{\partial^2 V}{\partial r'^2 } \left ( \frac{\partial  r'}{\partial p_0} \right )^2
 + \frac{\partial V}{\partial r' } \frac{\partial^2  r'}{\partial p_0^2}
 \right ] (p_0-p_{0\alpha }) . \label{kp-2p0} \ee
 Due to exponential divergence of neighboring trajectories in phase space,
 the main contribution to the right hand side of Eq.~(\ref{kp-2p0})
 comes from times $t' \approx t$.
 The time evolution of the quantity inside the bracket in Eq.~(\ref{kp-2p0}) is given by
 the dynamics of the system described by $H_0$.
 On average it increases as $ [\delta x(t)/ \delta x(0)]^2$, with $\delta x(t)$ being
 distance in phase space.
 With increasing time, the number of stationary points
 increases exponentially,
 roughly in the same way as  $ \delta x(t)/ \delta x(0)$,
 since the oscillation of $\Delta S$ is mainly induced by local instability
 of trajectories.

 Then, substituting Eq.~(\ref{kp-2p0}) into Eq.~(\ref{Mt-It}), we have
 \be \ov M(t) \propto \ov { \left [ \frac 1 {D|\delta x(t)/ \delta x(0)|} \right ]}. \label{Mt-dx1} \ee
 When $D$ changes slowly with $p_0$ and $\ww r_0$, we have
 \bey \nonumber  \ov M(t) \propto I_{\Lambda }(t) \equiv  e^{-\Lambda_1 (t) t}, \ \ \  \text{with}
 \\ \Lambda_1 (t)  = - \lim_{\delta x(0) \to 0} \frac 1t  \ln
 \overline{ \left |\frac{ \delta x(t)}{ \delta x(0)} \right |^{-1} }.
\label{It-3}  \eey
 In systems with constant local Lyapunov exponents,
 Eq.~(\ref{It-3}) reduces to the usual Lyapunov decay with $\Lambda_1 (t) =\lambda $.
 On the other hand, when fluctuation in local Lyapunov exponent cannot be neglected,
 $I_{\Lambda }(t)$ coincides with the $e^{-\lambda_1t}$ decay in Ref.~\cite{STB03}
 in the limit $t \to \infty $, with $\lambda _1 = \lim_{t \to \infty } \Lambda_1 (t) $.

 \subsection{Decay of the average of $\ln M(t)$}
 \label{sect:avg-ln}

\begin{figure}
\includegraphics[width=\columnwidth]{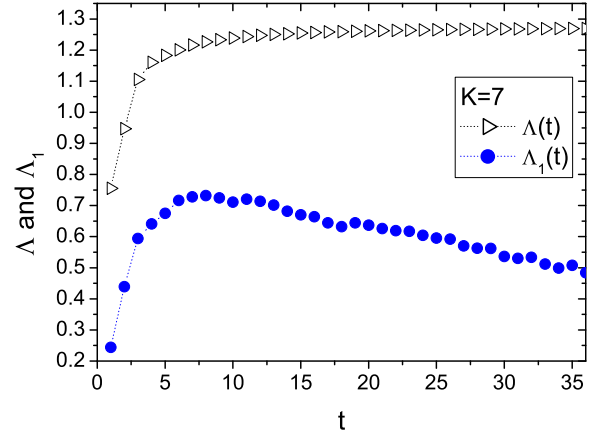}
 \caption{
 Variation of $\Lambda (t)$ and $\Lambda_1(t)$ with $t$ in the kicked rotator model at $K=7$,
 with average performed over $10^6$ random initial positions in phase space.
 The definitions of $\Lambda (t)$ and $\Lambda_1 (t)$ are given in Eqs.~(\ref{Lamb-t})
 and (\ref{It-3}), respectively.
 The value of $\Lambda (t)$ approaches the Lyapunov exponent $\lambda \approx 1.27 $ quickly, as $t$ increases.
 } \label{fig-lamb-t-k7}
\end{figure}

 To understand the decaying behavior of the average of $\ln M(t)$ for initial Gaussian
 wave packets, we divide the time $t$ into
 four time intervals, specifically, $(0,\tau_1), (\tau_1, \tau_2), [\tau_2, t_d)$, and $[t_d,t_s)$.
 Here $t_d$ is a time scale defined below, beyond which diagonal approximation can  be used
 before average is performed,
 and $t_s$ is the time at which the saturation value of fidelity is reached.

 Within the first time interval, $M(t)$ is described by Eq.~(\ref{Mt-gauss-short-t}) for most of initial states.
 For $\sigma $ not large, $\ln M(t)$ is close to zero.
 On the other hand, for quite large $\sigma $, the average of $\ln M(t)$ can be quite small
 (cf.~Fig.~\ref{fig-short-t-V1st}).

 For the second time interval $(\tau_1,\tau_2)$, Eq.~(\ref{M-skp}) can be used to calculate
 $\ln M(t)$, for initial states with $\ww k_p$ satisfying $|\sigma \ww k_p | \gg \pi /w_p $.
 For this part of initial states, $|\ww k_p|$ typically increases as
 $\delta x(t) / \delta x(0)$ [see Eq.~(\ref{partial-ds})], as a result,
 the average of $\ln M(t)$ for this part of initial states behaves as $-2\Lambda (t) t$,
 where $\Lambda (t)$ is defined by Eq.~(\ref{Lamb-t}).
 On the other hand, for initial states with small $|\ww k_p|$,
 which lie in the neighborhoods of the stationary points of $\Delta S$, fidelity decay is slower.
 Since the total size of the regions of $p_0$ with small $|k_p|$
 is small compared with the domain of $p_0$,
 the average of $\ln M(t)$ over all possible initial states has a decay rate between
 $\Lambda (t)$ and $2\Lambda (t)$.
 In a classical system with strong chaos, $\Lambda (t)$ approaches the
 Lyapunov exponent $\lambda $ quickly, as seen in Fig.~\ref{fig-lamb-t-k7},
 therefore, in this time interval, the decay rate of the average of $\ln M(t)$ is usually close to, or a little
 larger than $\lambda $.

 For $t$ around $\tau_2$ or longer, the main contribution to the average of $\ln M(t)$
 is given by the neighborhood of stationary points and
 one can start from Eq.~(\ref{msc-ma}) in calculating $M(t)$.
 In the third time interval $[\tau_2, t_d)$, the number of stationary points within the effective window in $p_0$ is
 small, hence, diagonal approximation cannot be used in calculating the absolute value square of
 the right hand side of Eq.~(\ref{msc-ma}).
 We do not know much about the decay of the average of $\ln M(t)$ in this time interval.
 One should note that the third time interval is quite short,
 as a result of the exponential increment of the number of oscillations of $\Delta S$.

 In the fourth time interval $[t_d,t_s)$, diagonal approximation is applicable
 to the absolute value square of the right hand side of Eq.~(\ref{msc-ma}) before average is performed,
 due to the large number of stationary points within the effective window in $p_0$.
 Then, Eqs.~(\ref{msc-ma}) and (\ref{ma-st}) give
 \bey \nonumber  M(t) & \simeq & { \sum_{\alpha } |m_{\alpha }(t)|^2 }
 \\ & = & \sum_{\alpha} \frac{2\hbar }{w_p^2 }
 \frac{ {\exp } \left [ - 2(p_{0\alpha } - \ww p_0 )^2/w_p^2 \right ] }
 {{ | \Delta {S_{\alpha }^{''}} | }}.   \eey
 Using Eq.~(\ref{partial-ds}), we write $\Delta S_{\alpha }'' $ as
 \be \Delta S_{\alpha }''  \approx \epsilon \int_0^t dt' \left [
 \frac{\partial^2 V}{\partial r'^2 } \left ( \frac{\partial  r'}{\partial p_0} \right )^2
 + \frac{\partial V}{\partial r' } \frac{\partial^2  r'}{\partial p_0^2}
 \right ] . \label{S''} \ee
 Arguments similar to those leading from Eq.~(\ref{kp-2p0}) to Eq.~(\ref{Mt-dx1}) show that,
 when the fluctuation of $|\delta x(t)/ \delta x(0)|$ is small for
 $p_0$ within the effective window in the $p_0$ space,
 the main decaying behavior of a single fidelity is typically
 \be M(t) \propto  |\delta x(t)/ \delta x(0)|^{-1}, \label{Mt-dx2} \ee
 with $\delta x(0)$ being a small displacement from $(\ww r_0, \ww p_0)$, the center of the initial Gaussian packet.
 Then, it is ready to obtain
 \be \exp [ \ \ov {\ln M(t) } \ ] \propto e^{ -\Lambda (t) t} \simeq e^{ -\lambda t}.  \label{Mt-lyap} \ee
 where the second equation is obtained, since
 $\Lambda (t)$ is usually close to  the Lyapunov exponent $\lambda $ within this time interval.

 Since the fluctuation of $|\delta x(t)/ \delta x(0)|$ within the effective window in $p_0$ increases with time,
 Eq.~(\ref{Mt-lyap}) becomes invalid for $t$ long enough.
 In the long time limit,
 the fluctuation of $|\delta x(t)/ \delta x(0)|$ within the effective window in $p_0$ has
 similar properties as in the whole $p_0$ domain, then, the average of $\ln M(t)$
 has the $\Lambda_1(t)$ decay in Eq.~(\ref{It-3}), with $\Lambda_1(t)$ taking the value $\lambda_1$.
 For intermediate times, it is reasonable to expect that the decaying rate of
 the average of $\ln M(t)$ decreases from the Lyapunov exponent and approaches $\Lambda_1(t)$
 with increasing $t$.

 Combining the above results, it is seen that there indeed exists a certain short time interval,
 in which the average of $\ln M(t)$ follows roughly the Lyapunov decay,
 as observed in the kicked top model \cite{WL02}.
 Specifically, the decaying rate of the average of $\ln M(t)$ is close to or a little larger than the Lyapunov
 exponent in the second time interval;
 then, beyond the short third time interval, it decreases
 from the Lyapunov exponent and approaches $\Lambda_1 (t)$ at long time.

\begin{figure}
\includegraphics[width=\columnwidth]{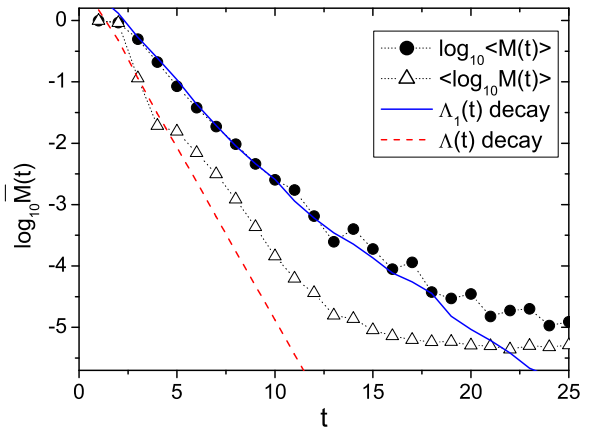}
 \caption{
 Fidelity decay in the kicked rotator model at $K=7$, $N=2^{17}$, $\sigma =20$.
 Average is performed over 500 initial Gaussian wave packets with $\xi = \sqrt{\hbar }$,
 whose centers are chosen randomly in the chaotic sea in phase space.
 The $\Lambda (t)$ and $\Lambda_1(t)$ decays are the predictions in
 Eqs.~(\ref{Mt-lyap}) and (\ref{It-3}), respectively,
 with the values of $\Lambda (t)$ and $\Lambda_1(t)$ shown in Fig.~\ref{fig-lamb-t-k7}.
 Decay of the average of $M(t)$ is described by the $\Lambda_1(t)$ decay.
 The average of $\ln M(t)$ is a little faster than the $\Lambda (t)$ decay in an initial
 short time interval, as predicted by the theory.
 } \label{fig-Mt-kr-k7}
\end{figure}

 We have tested these predictions, as well as those in the previous section,  in the kicked rotator model
 at $K=7$, which are shown in Fig.~\ref{fig-Mt-kr-k7}.
 It is seen in the figure that the average of $\ln M(t)$ decays a little faster than the
 $\Lambda (t)$ decay (as well as the Lyapunov decay) initially,
 then, after a transient time, it decays a little slower than the Lyapunov decay,
 but obviously faster than the $\Lambda_1(t)$ decay.
 The predicted long time $\Lambda_1(t)$ decay for the average of $\ln M(t)$ is not seen at $K=7$,
 because $K$ is not large enough for the value of $N$ taken.
 We mention that the $\lambda_1$ decay at long time has indeed been observed for $K=10$ in Ref.~\cite{STB03}.

 Finally, for quite large $\sigma $, the initial approximate Lyapunov decay discussed above may disappear.
 Indeed, in this case,  $|\ln M(t)|$ can be quite large for $t<\tau_1$, as seen in Eq.~(\ref{Mt-gauss-short-t}).
 The average of $\ln M(t)$ for short time $t$ is in fact dominated by contributions described by
 Eq.~(\ref{Mt-gauss-short-t}), even beyond $\ov {\tau}_1$,
 since the value of $\tau_1$ fluctuates with respect to the initial condition,
 i.e., for some initial states $\tau_1$ can be relatively long.
 Hence, the average of $\ln M(t)$ can be obviously smaller than the prediction of
 the Lyapunov decay in the second time interval, as shown numerically in \cite{STB03}.

 \section{Conclusions and discussions}
 \label{sect:con}

 In this paper, we have improved the uniform semiclassical approximation to fidelity
 by considering the second order term in the Taylor expansion of action,
 which is important for fidelity of initial Gaussian wave packets with width of the order $\sqrt {\hbar }$.
 Short-time-decay of fidelity is analyzed, which is initial-state-dependent,
 in particular, two time scales have been introduced and studied in detail for initial Gaussian wave packets.
 Initial FGR decay of fidelity in the perturbative regime is
 confirmed by direct numerical calculations.
 Non-FGR decay in the FGR regime in systems with weak chaos in the classical limit
 is explained by relating the distribution of action difference to the  Levy distribution.
 The average of the logarithm of fidelity is shown to have an approximate Lyapunov decay within some time intervals,
 in the Lyapunov regime in systems possessing large fluctuations of the finite-time Lyapunov exponents
 in the classical limit.

\begin{figure}
 \includegraphics[width=\columnwidth]{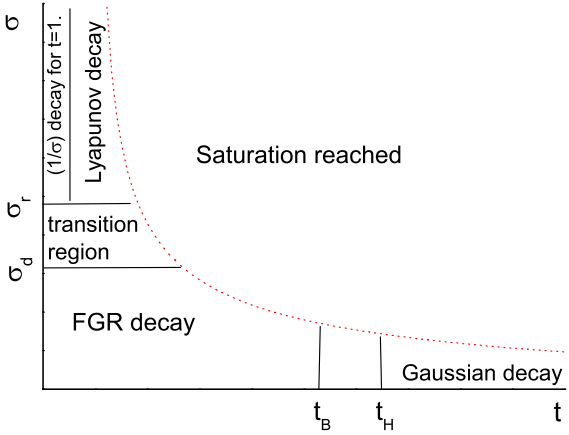}
 \caption{
 Schematic diagram for fidelity decay of initial point sources
 in systems with constant local Lyapunov exponents in the classical limit.
 $t_B$ is the breakdown time of semiclassical approximation,
 and $t_H $ is the Heisenberg time, beyond which the Gaussian decay sets in.
 $\sigma_d$ and $\sigma_r$ are two perturbation scales introduced in Ref.~\cite{WCL04}.
 Below $\sigma_d$, one has the FGR decay for $t<t_B$,
 and above $\sigma_r$ is the Lyapunov regime, in which the Lyapunov decay appears beyond the first kick.
 } \label{fig-phase-d-st-point}
 \end{figure}

 As we have demonstrated that
 fidelity has a decaying behavior richer than the simple picture mentioned in the
 beginning of the introduction with just four distinct regimes.
 In Fig.~\ref{fig-phase-d-st-point}, we present a schematic diagram for the present
 understanding of fidelity decay of initial point sources in systems
 possessing constant local Lyapunov exponent in the classical limit.
 For initial Gaussian wave packets, short time decay is more complex than for initial point sources,
 for which two time scales $\tau_1$ and $\tau_2$ should be introduced, as discussed in this paper.
 When the underlying classical dynamics has large fluctuation in
 the finite-time Lyapunov exponent, the decaying rate of fidelity in the
 Lyapunov regime is not given by the Lyapunov exponent.

 \acknowledgments

 The work was supported in part by a Faculty Research Grant of National University of Singapore,
 the Temasek Young Investigator Award of DSTA Singapore under Project Agreement POD0410553 (BL),
 and the Natural Science Foundation of China No.10275011 (WGW).

 \appendix

 \section{Dependence of fidelity Decay on perturbation strength in deep Lyapunov regime for initial point sources}
 \label{sect:proof-1st-kick}

 In this appendix, we consider the dependence of fidelity decay on perturbation strength in the
 deep Lyapunov regime for initial point sources, which can be estimated by using Eq.~(\ref{mt-point-p0}).
 Let us first divide the domain $[0,2\pi )$ of $p_0$ into a series of subregions,
 $ [p_{0j},p_{0(j+1)} )$, so that within each of the subregions the phase $\Delta S / \hbar $
 of the integrand on the right hand side of Eq.~(\ref{mt-point-p0}) can be approximated by a linear function,
 \be \Delta S / \hbar \simeq  \sigma (k_j p_0 + b_j), \hspace{0.2cm} \text{for} \ p_0 \in [p_{0j},p_{0(j+1)} ),
 \label{gp0} \ee
 where the parameters $k_j$ and $b_j$ do not depend on $p_0$.
 Here $p_{0j}=0$ for $j=1$ and $p_{0j}=2\pi $ for  $j=N_X+1$,
 with $N_X$ denoting the number of the subregions.
 Substituting Eq.~(\ref{gp0}) into Eq.~(\ref{mt-point-p0}), we have
 \bey \nonumber \label{mt-point-gp0} m_{p }(t) \simeq \frac 1{2 \pi i  \sigma } \sum_{j=1}^{N_X} X_j
 \hspace{1cm} { \rm with }
 \\ X_j := \frac 1{k_j} \left ( e^{i \sigma k_j p_{0(j+1)}}
 -  e^{i \sigma k_j p_{0j}} \right ) e^{i \sigma b_j }. \label{Xj} \eey

 In arranging the subregions, we require that $N_X$ should be as small as possible,
 conditional on that the linear approximation in Eq.~(\ref{gp0}) does not lose the main contribution
 to the right hand side of Eq.~(\ref{mt-point-p0}).
 Since the phase $\Delta S / \hbar $ is proportional $\sigma $,
 $N_X$ increases approximately linearly with $\sigma $.
 In the deep Lyapunov regime with quite large $\sigma $, for the subregions chosen in this way,
 the phase of $X_j$, which is approximately proportional to $\sigma $, can be regarded as random
 with respect to $j$ and $r_0$,
 then, diagonal approximation can be used in computing the averaged fidelity,
 \be \label{ov-M} \overline M_p(t) =  \frac 1{2 \pi } \int dr_0 |m_p(t)|^2
 \simeq \frac 1{(2 \pi \sigma )^2}  \sum_{j=1}^{N_X} \ov { | X_j|^2 }, \ee
 where we assume that fluctuation of $N_X$ with respect to $r_0$ is small.
 Finally, using $N_X \propto \sigma $, we have the following approximation,
 \be \ov M_p(t) \propto 1 / \sigma , \ \ \ \ \text{for} \ \sigma \gg 1 . \label{Mt-p-tau} \ee

 \end{document}